\documentclass[pdflatex,sn-basic]{sn-jnl}

\usepackage{graphicx}%
\usepackage{multirow}%
\usepackage{amsmath,amssymb,amsfonts}%
\usepackage{amsthm}%
\usepackage{mathrsfs}%
\usepackage[title]{appendix}%
\usepackage{xcolor}%
\usepackage{textcomp}%
\usepackage{manyfoot}%
\usepackage{booktabs}%
\usepackage{algorithm}%
\usepackage{algorithmicx}%
\usepackage{algpseudocode}%
\usepackage{listings}%

\theoremstyle{thmstyleone}%
\theoremstyle{thmstyletwo}%

\theoremstyle{thmstylethree}%

\begin{document}

\title[Improving Spatial Resolution of Background Oriented Schlieren Based on Directional Rays]{Improving Spatial Resolution of Background Oriented Schlieren Based on Directional Rays}


\author[1]{\fnm{Xiang} \sur{Li}} 
\equalcont{These authors contributed equally to this work.}

\author[1]{\fnm{Muen} \sur{Gao}} 
\equalcont{These authors contributed equally to this work.}

\author[1]{\fnm{Weiran} \sur{Wang}} 

\author[2]{\fnm{Jiawei} \sur{Li}} 

\author[1]{\fnm{Chong} \sur{Pan}} 

\author[1]{\fnm{Jinjun} \sur{Wang}} 

\author*[1,2]{\fnm{Yuan} \sur{Xiong}}\email{xiongyuan@buaa.edu.cn}

\affil*[1]{\orgdiv{Institute of Fluid Mechanics}, \orgname{Beihang University}, \orgaddress{\street{37, Xueyuan Road}, \city{HaiDian District}, \postcode{100191}, \state{Beijing}, \country{China}}}

\affil[2]{\orgdiv{Ningbo Institute of Technology}, \orgname{Beihang University}, \orgaddress{\street{399 Kangda Road}, \city{Ningbo}, \postcode{315800}, \state{Zhejiang}, \country{China}}}



\abstract{The background-oriented Schlieren technique has emerged as a promising method for visualizing density gradients and performing quantitative measurements. However, an inherent constraint of BOS is the compromise between spatial resolution and measurement sensitivity, as the BOS camera typically remains focused on the background pattern. To overcome the resolution-sensitivity constraint, a new variant of BOS based on nominally directional rays has been proposed in this paper. Instead of utilizing diffusively reflective background patterns, a spherically concave mirror etched with random dots has been used to create a dotted background that reflects rays directionally. Combined with coaxial LED light illumination, we demonstrate that the current setup can improve the spatial resolution of canonical BOS without compromising measurement sensitivity. Moreover, the proposed setup decouples the requirement of a small lens aperture to achieve a large depth of field, thereby significantly alleviating the need for strong background light illumination in high-speed BOS applications. To demonstrate the effectiveness of the proposed method in improving the BOS spatial resolution, both synthetic BOS image generations and experiments on low- and high-speed jets are conducted. Results show that the proposed variant of BOS can be advantageous for measuring density-varying flows with a limited field of view.}




\maketitle

\section{Introduction}\label{intro}

Background-oriented Schlieren (BOS) has gained popularity in recent decades for visualizing and performing quantitative measurements over density gradients, owing to its advantages in easy calibration, flexible field of view, and simplified optical alignment as summarized by \cite{raffel_background-oriented_2015, settles_review_2017, schmidt_twenty-five_2025}. The canonical BOS optical setup consists of only a camera, an endocentric lens, and a patterned background, which typically reflects light diffusively. By comparing the background images with and without the target flow (phase object), quantitative background distortion displacements can be accurately extracted using proper algorithms, such as cross-correlations by \cite{richard_principle_2001, elsinga_assessment_2004, goldhahn_background_2007}. Practically, to ensure accurate distortion extraction, the BOS camera primarily focuses on the background pattern, leaving the phase object out of focus. Combined with the algorithmic window size required by displacement extraction algorithms, the spatial resolution of BOS is inevitably compromised and has been identified as an inherent challenge to canonical BOS measurements since the birth of the BOS method, as illustrated by \cite{elsinga_assessment_2004,goldhahn_background_2007,hargather_comparison_2012,gojani_measurement_2013}.

Significant progress has been made in eliminating the resolution compromise caused by the algorithm window. Within the cross-correlation framework, \cite{sourgen_reconstruction_2012} applied gliding interrogation windows to increase the quantity of displacement vectors with backgrounds filled with colorful dots. To achieve pixelwise dense displacement fields, the optical flow algorithm, initially introduced by the computer vision community, has been applied for BOS measurements by \cite{atcheson_evaluation_2009, grauer_fast_2020, schmidt_wavelet-based_2021, li_three-dimensional_2024}. To improve the displacement extraction efficiency while retaining the pixel-wise resolution, \cite{wildeman_real-time_2018} proposed the fast Fourier demodulation method, and later further developed by \cite{shimazaki_background_2022}. To cope with large density gradients, \cite{vinnichenko_performance_2023} found that Fourier transform profilometry can outperform other algorithms with respect to large displacement gradients. Recently, \cite{ota_spatial_2023} utilized a deep super-resolution network in achieving even a super-resolved displacement field compared to the pixel-wise resolution of optical flow algorithms. However, regardless of the density of the displacement vectors extracted via the algorithm, aperture-determined cone-shaped light rays still compromise the BOS spatial resolution due to the hardware configuration. 

Efforts to reduce the light-cone reduced spatial resolution of BOS have initially focused on optimizing the canonical BOS setup parameters without modifying the hardware configuration. For example, it is a common practice to position the phase object close to the background to keep both the background and the phase object within the depth of field of the camera, thereby improving the spatial resolution as demonstrated by \cite{bichal_application_2014, nicolas_3d_2017}. While such an approach could minimize the blurriness of the phase object, \cite{gojani_measurement_2013} pointed out that the displacement reduces its amplitude as the phase object moves closer to the background due to a lower measurement sensitivity. Further considering that the BOS field of view (FOV) needs to remain constant to fully utilize the sensor resolution, \cite{lang_measurement_2017} reformulated the expression for the BOS sensitivity and found that the maximum sensitivity is always achieved by allocating the phase object in the middle between the camera and the background. To balance the spatial resolution and measurement sensitivity, \cite{schwarz_practical_2023} suggested that once the minimal sensitivity requirement is satisfied, the parameters can be adjusted to maximize the spatial resolution of BOS. However, for a large test section, placing both the flow and background within the FOV still becomes difficult even with a minimized lens aperture together with an ultra-strong background illumination from over-driven LEDs or lasers, as shown by \cite{nicolas_3d_2017, xiong_towards_2020}.  

Instead of optimizing the canonical BOS setup parameters, the light-cone decreased spatial resolution can also be improved by modifying the BOS hardware configuration. For example, \cite{elsinga_assessment_2004} and \cite{ota_improvement_2015} added additional lenses to the BOS setup to enable Schlieren- or telecentric-mode of BOS, which can improve the BOS spatial resolution by collecting parallel projections of the background through the test section. \cite{hirose_double-pass_2023} has further enhanced the telecentric BOS sensitivity by enabling dual light paths, while retaining the high spatial resolution feature. However, to select those rays parallel to the optical axis with a telecentric setup, a small-sized aperture is still required to guarantee a sufficient depth-of-field, thus limiting the image brightness and contrast. In addition to the telecentric BOS, \cite{meier_improved_2013} proposed a laser speckle BOS technique with several operational variants. When operating in interference mode, the camera can focus directly on the flow with adjustable sensitivity. However, laser speckle dimension relies on the aperture size, and to avoid peak-locking effects, a relatively small aperture size is again required. Moreover, the laser speckle is highly sensitive to the status of the laser oscillator and the speckle screen, making this approach challenging to apply in harsh environments, such as high-speed wind tunnels. 

Recently, several attempts have been made to incorporate neural networks into BOS reconstructions by \cite{molnar_forward_2024, li_neural_2024, he_neural_2025, molnar_algorithm_2025,molnar_open-source_2025}. Thanks to neural networks, the BOS spatial resolution can be boosted without requiring modifications to the hardware. Inspired by the photography, the defocus blurring is caused by the averaging across the cone-shaped light ray bundles. \cite{schwarz_practical_2023,molnar_forward_2024} assumed a Gaussian or disk filter as the forward model for the defocus blurring in BOS. \cite{molnar_forward_2024} successfully improved the spatial resolution of a supersonic flow by incorporating a cone-ray defocus blurring model into a neural-implicit reconstruction framework. However, considering that physics-informed neural networks are usually time-consuming due to the optimization process, the feasibility of enhancing the spatial resolution within the canonical BOS reconstruction framework still requires further exploration.

In this paper, we propose a new variant of BOS termed the directional-ray BOS (DRBOS), which significantly reduces the cone-shaped light-induced compromise in the spatial resolution. This approach eliminates the need for a small aperture size in the canonical BOS setup, thereby alleviating the requirement for strong background illumination. Moreover, the measurement sensitivity also has no compromise compared to the canonical setup. The rest of the paper is organized as follows: Sec. \ref{sec_Methodology} introduces the principle of canonical BOS and DRBOS; Sec. \ref{sec:DRBOS_Res_Sens} and Sec. \ref{sec:DRBOS_dualpath_error} present the derivation of the sensitivity expression for the DRBOS and associated validation with synthetic images; Sec. \ref{sec:he_jet_exp} and Sec. \ref{sec:supersonic_jet_exp} further verify the spatial resolution superiority of DRBOS with both sub- and supersonic jet experiments; Sec. \ref{sec:Conclusions} provides the conclusion of this study.

\section{Methodology}\label{sec_Methodology}
\subsection{BOS principle}\label{sub_endocentric_BOS}
\begin{figure}[!htb]
  \centering
  \includegraphics[width=70mm]{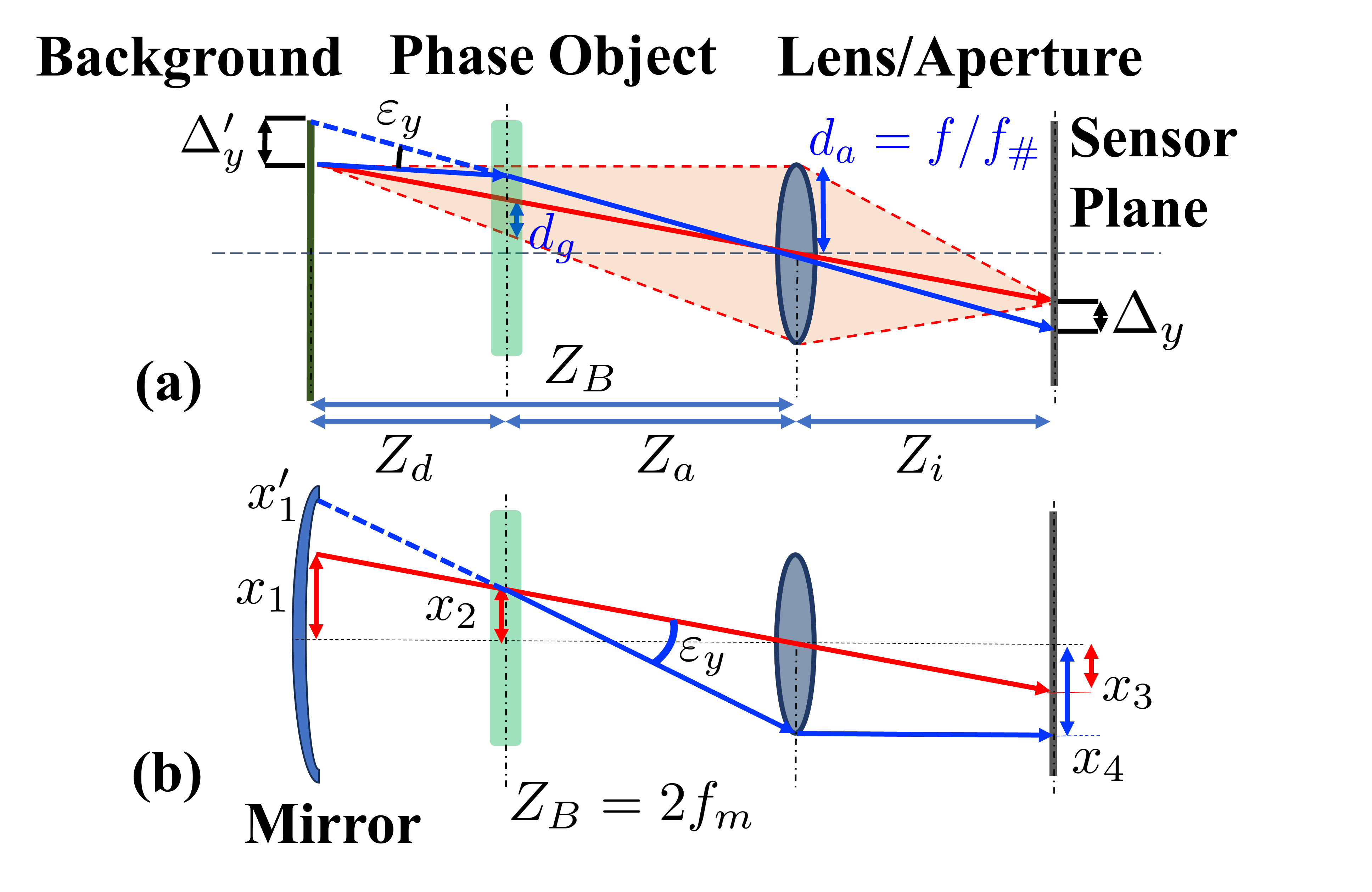}
\caption{A schematic for the endocentric BOS system (a) and for the DRBOS system (b).}
\label{fig:BOS schematic} 
\end{figure}

A canonical endocentric BOS setup typically consists of a camera, a background with random dots, and a light source for illuminating the background, as shown schematically in Fig.\ref{fig:BOS schematic}(a). 
The target flow of a non-uniform refractive index ($n$) is allocated between the camera lens and the background. The ray equation governs the trajectory of a light ray within the phase object:
\begin{equation}
\frac{d}{ds}(n \frac{d\mathbf{r}}{ds})=\nabla{n}
\label{eq_RayEq}
\end{equation}
where $\mathbf{r}$ is the position vector along the ray path, $ds$ denotes the infinitesimal path element, and $d\mathbf{r}/ds={\mathbf{d}}$ represents the unit direction vector of the ray. For a two-dimensional phase object, the deflection angle of the light ray $\varepsilon_y$ shown in Fig.\ref{fig:BOS schematic}(a) is typically approximated as:
\begin{equation}
\varepsilon_y \approx \Delta_y^\prime/Z_d.
\label{eq_deflection_geometric}
\end{equation}
Inside, the displacements $\Delta_y^\prime$ on the image plane were typically calculated by applying the multiple-pass cross-correlation algorithm to the image pair with the random dot pattern. Consider the integrated Eq.~(\ref{eq_RayEq}) with paraxial assumptions, $\varepsilon_y$ can be linked to $n$ as:
\begin{equation}
\varepsilon_y\approx\frac{1}{n_0}\int_S \frac{\partial n}{\partial y} dz
\label{eq_epsilonDefine}
\end{equation}
In the two-dimensional case, considering both $\varepsilon_x$ and $\varepsilon_y$ results in tackling the BOS problem equivalent to solving a Poisson equation with proper boundary conditions as shown by \cite{xiong_analysis_2020}. Once $n$ field is obtained, owing to the Gladstone–Dale (G-D) relation:
\begin{equation}
n-1=G\rho,
\label{eq_gdrelation}
\end{equation}
the flow density $\rho$ can be inferred from $n$ provided the G-D constant $G$ which is determined by the light wavelength and gas composition. More details regarding the canonical BOS methodology can be found in the aforementioned BOS reviews.

\subsection{DRBOS principle}\label{sub_DRBOS}

An inherent feature of canonical BOS lies in the utilization of an endocentric lens to collect diffusely reflected light from the background. Taking the reference rays for example, instead of a single ray from the background (the red solid line in Fig.\ref{fig:BOS schematic}(a)), a bundle of rays within a light cone transverse through the phase object as the shaded area. The opening angle of the cone is controlled by the dimension of the lens aperture $d_a$, defined as the division of the lens focal length $f$ by the aperture number $f_\#$. According to a simple geometric relationship, the projected dimension of the light ray cone on the phase object is $d_g=Z_dd_a/Z_B$. Physically, it means all the deflection information carried by the light rays within $d_g$ is averaged into a single displacement $\Delta_y^\prime$. Such an average effect is the well-known 'defocus blurriness' in BOS, which compromises the spatial resolution of BOS measurements.

\begin{figure}[!htb]
  \centering
  \includegraphics[width=67mm]{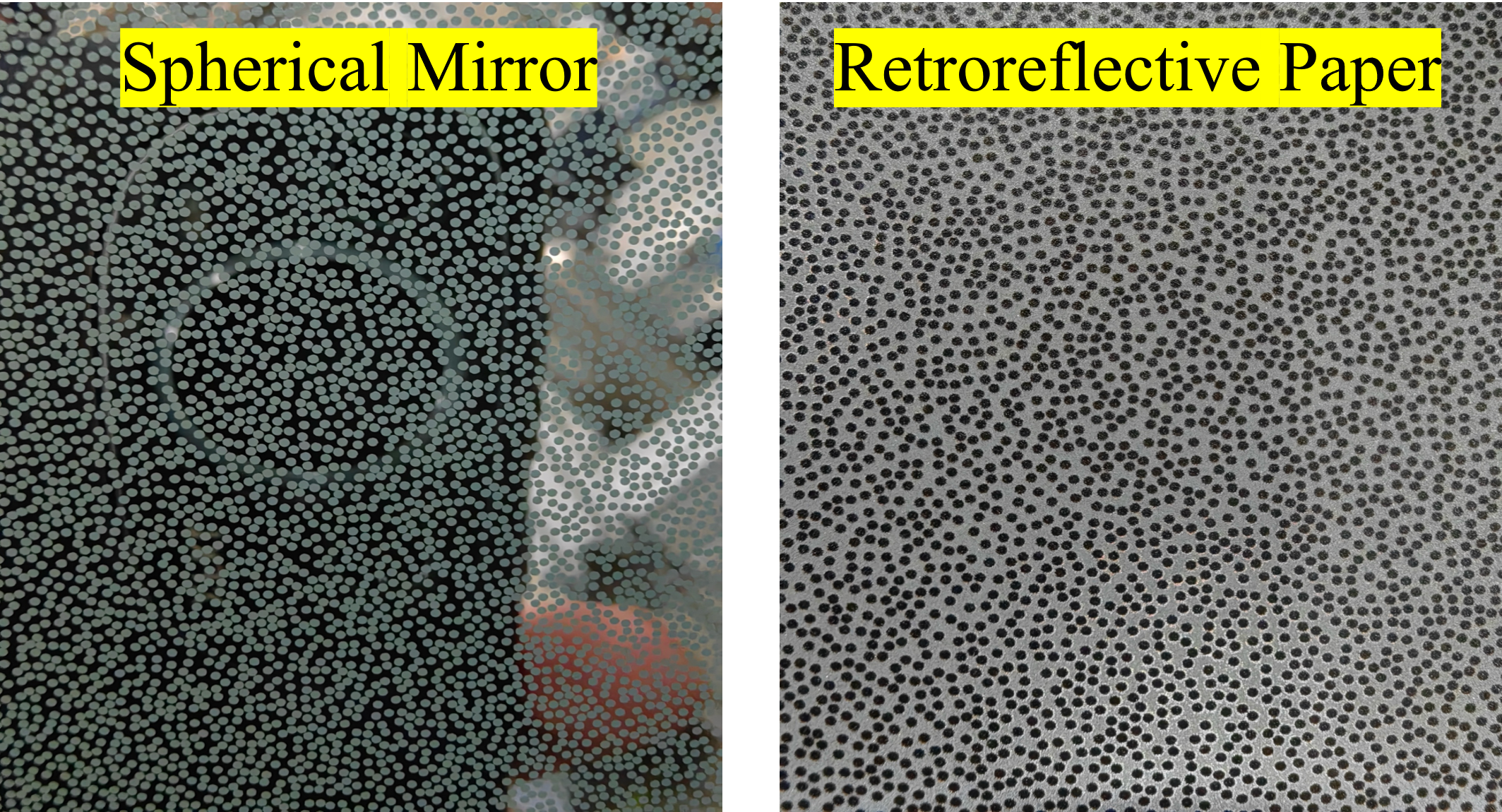}
\caption{Comparison between the SCM and retroreflective background, both with random dots.}
\label{fig:random_dots}       
\end{figure}

A typical schematic of DRBOS is shown in the Fig.\ref{fig:BOS schematic}(b). The essential feature of DRBOS is to have nominally one ray reflected from one point on the background, without forming the ray bundle controlled by the lens aperture as in the canonical BOS. To achieve directional rays with a background printed with random dots, a spherical concave mirror (SCM) etched with random dots is utilized. Directional rays are produced according to the feature of the SCM that rays emitted at the $2f_m$ along the optical axis will return to the same position after the SCM's reflection. Inside $f_m$ is the focal length of the SCM. A convex lens etched with random dots is expected to function similarly when receiving parallel rays. Note that the ray concept falls within the category of geometric optics, thus the diffraction interferences are not considered here. To ensure the etching quality of the random dot, an infrared laser is utilized to remove the reflective coating of the mirror precisely. 

A comparison between etched SCM and the retroreflective paper printed with the same pattern is shown in Fig.\ref{fig:random_dots}. The superior quality of the random dots achieved by the etching approach can be confirmed. Compared to the canonical BOS, the light deflection for DRBOS is distinct from the canonical one. As shown in Fig.\ref{fig:BOS schematic}, in BOS the deflected rays (blue solid line) differ from the reference ray (red solid line) between the phase object and the background. As a comparison, the light path remains the same in DRBOS owing to the SCM. Nevertheless, the deflection angle estimations for DRBOS follow the same equation (Eq.\ref{eq_deflection_geometric}). Note that to simplify the ray path for DRBOS, the DRBOS background is assumed to emit the directional ray here and the associated error is explored in section \ref{sec:DRBOS_dualpath_error}. To examine the feasibility of DRBOS and quantify the associated measurement error, we utilized synthetic BOS image generation based on the nonlinear ray tracing, details of which can be found in recent papers by \cite{rajendran_pivbos_2019,jia_tomographic_2024,li_neural_2024,li_three-dimensional_2024}. Flow fields involved for creating the synthetic images are two-dimensional slices of the buoyancy-driven turbulence from the Johns Hopkins Turbulence Database, as shown by \cite{li_public_2008}. Displacements are calculated precisely based on the weight centre shift of ray bundles on the image sensor with and without the flow field, excluding the errors introduced by the displacement extracting algorithm. Detailed settings for the synthetic image generation are shown in Table \ref{tab:synthetic-JHU}.

\begin{table}[!htb]
\caption{Parameters for BOS synthetic image generation with the buoyancy-driven turbulent flows.}\label{tab:synthetic-JHU}%
\begin{tabular}{@{}llll@{}}
\toprule
Parameter & Setting \\
\midrule
$Z_{B}$ &  $1500$ mm \\
$Z_{d}$ & 750 mm \\
Voxel Res.  & $500 \times 500 \times 10$ voxels \\
Volume Size   & $50 \times 50 \times 1$ mm$^3$ \\ 
$f_{\mathrm{Lens}}$ & $50$ mm \\
$f_{\#}$ & $22$, $8$, $4$ \\
Image resolution & $1000 \times 1000$ pixels\\
Pixel Pitch & $3.45 \times 3.45$ $\mu$m$^2$\\
Reference density field & $1.225$ kg/m$^3$ \\
Gladstone-Dale constant & $2.25 \times 10^{-4}$ m$^3$/kg \\
\botrule
\end{tabular}
\end{table}

\begin{figure}[!htb]
  \centering
  \includegraphics[width=120mm]{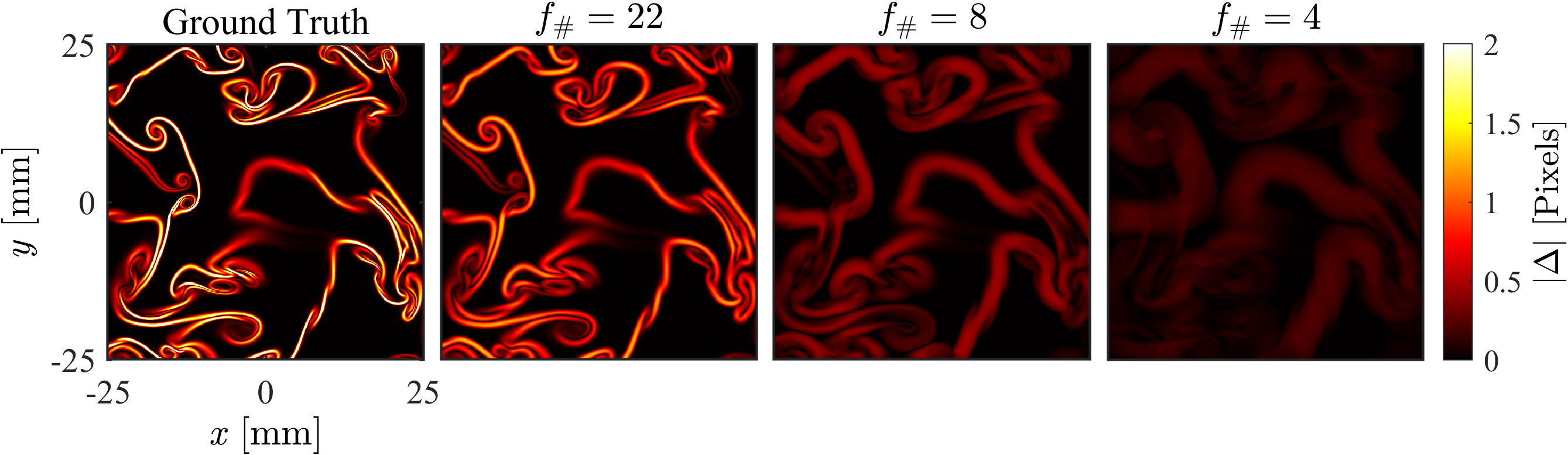}
\caption{Dependency of the displacement fields over the aperture number $f_{\#}$ for canonical BOS with the parameter setting in Table \ref{tab:synthetic-JHU}}
\label{fig:syn_BOS_fhash}       
\end{figure}

The synthetic image generation platform is tested first by exploring the dependency of the spatial resolution of the canonical BOS on the lens aperture $f_{\#}$. Associated displacement fields with different $f_{\#}$ are compared to the DNS ground truth in Fig. \ref{fig:syn_BOS_fhash}. The blurriness of the displacement field is obviously proportional to the value of $f_{\#}$ due to the increased opening angle of cone-shaped light rays. These observations are consistent with previous BOS experimental data reported in the literature, thus confirming the effectiveness of the synthetic image generation approach utilized here.

\subsection{Experimental Setup}\label{sec:sub_expsetup}
\begin{figure}[!htb]
  \centering
  \includegraphics[width=75mm]{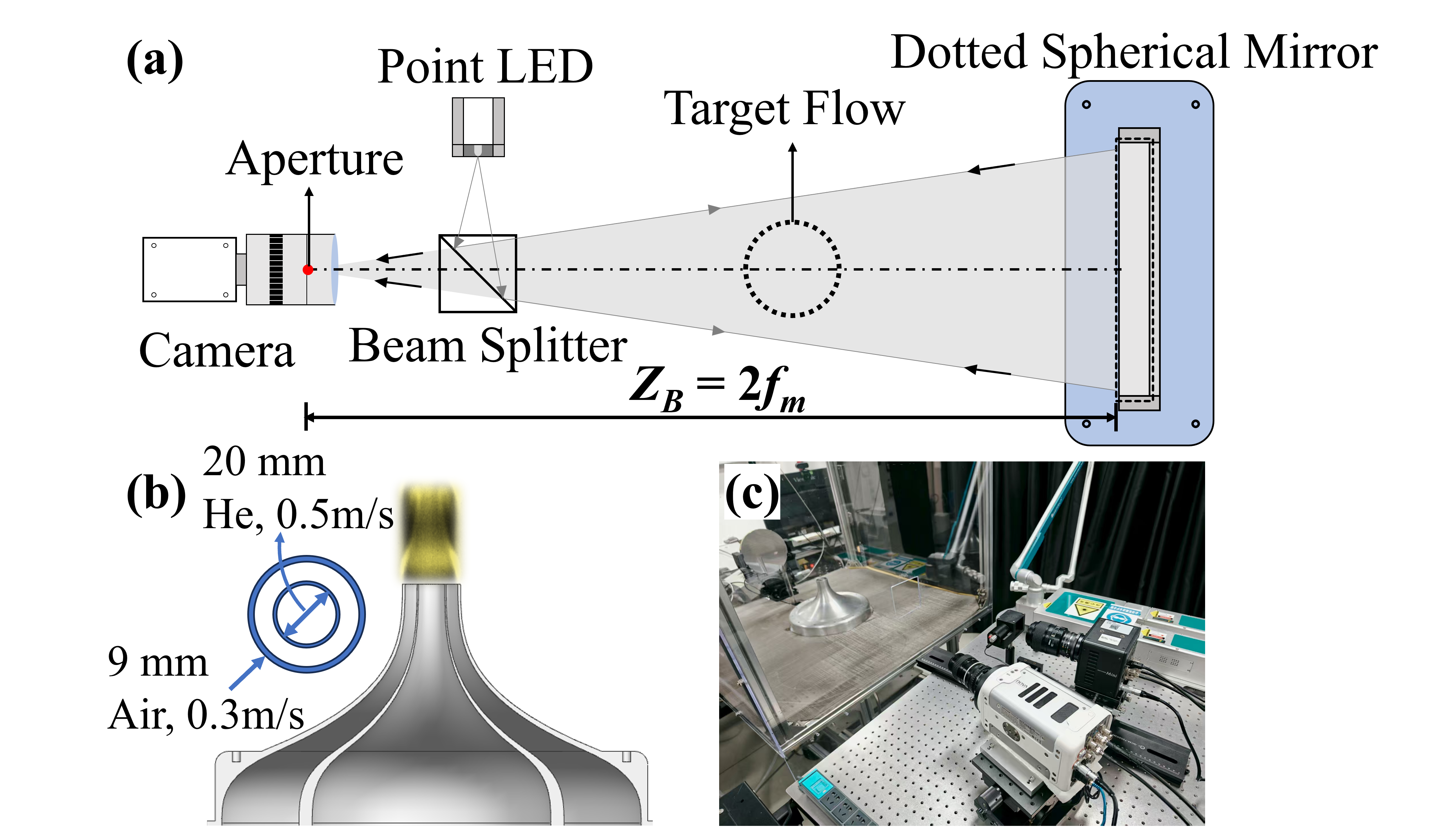}
\caption{The schematic for the Helium jet experiments with DRBOS (a); the contraction nozzles with coflow air to generate the Helium jet with top-hap outlet profile (b); experimental setup for DRBOS-PIV with Helium jet (c).}
\label{fig:DRBOS_helium_schematic}       
\end{figure}

In addition to the synthetic image tests, two laboratory experiments — utilizing a subsonic Helium jet and a supersonic underexpanded air jet — are employed to evaluate the performance of DRBOS. For the Helium jet, flow density variations are caused by the mixing between the Helium and the surrounding air due to the shear layer development and the diffusion process. The schematic of the Helium jet setup is shown in Fig.\ref{fig:DRBOS_helium_schematic}(a). A coaxial configuration of the LED and the camera is utilized, allowing the light ray to pass through the phase object twice before reaching the sensor plane. The distance between the lens aperture and the mirror is set to $2f_m$. To generate a Helium jet with a top-hat outlet velocity profile, the contraction nozzle with a bicubic curve is designed with air coflow to protect the Helium flow as illustrated by \cite{huang_physics-augmented_2025} as shown in Fig.\ref{fig:DRBOS_helium_schematic}(b). The described DRBOS setup together with the PIV system is demonstrated in Fig.\ref{fig:DRBOS_helium_schematic}(c) with the detailed system setting parameters shown in Table \ref{tab:helium jet}.

\begin{table}[!htb]
\caption{System parameters for DRBOS and BOS with the subsonic Helium jet.}\label{tab:helium jet}%
\begin{tabular}{@{}llll@{}}
\toprule
Parameter & Setting \\
\midrule
SCM diameter(DRBOS) &  $203$ mm \\
Helium jet diameter &  $20$ mm \\
Air coflow diameter &  $40$ mm \\
$f_{\mathrm{m}}$ & $750$ mm \\
$Z_{B}$ &  $1500$ mm \\
$Z_{d}$ & 450 mm \\
Image resolution & $1280 \times 1024$ pixels\\
Pixel size & $10 \times 10$ $\mu$m$^2$\\
Frame rate & $1000$ Hz\\
Exposure time & $250$ $\mu$s\\
$f_{\mathrm{Lens}}$ & $105$ mm \\
$f_{\#}$ & $2.8$, $5.6$, $8.0$, $11$, $16$ \\
\botrule
\end{tabular}
\end{table}

\begin{figure}[!htb]
  \centering
  \includegraphics[width=67mm]{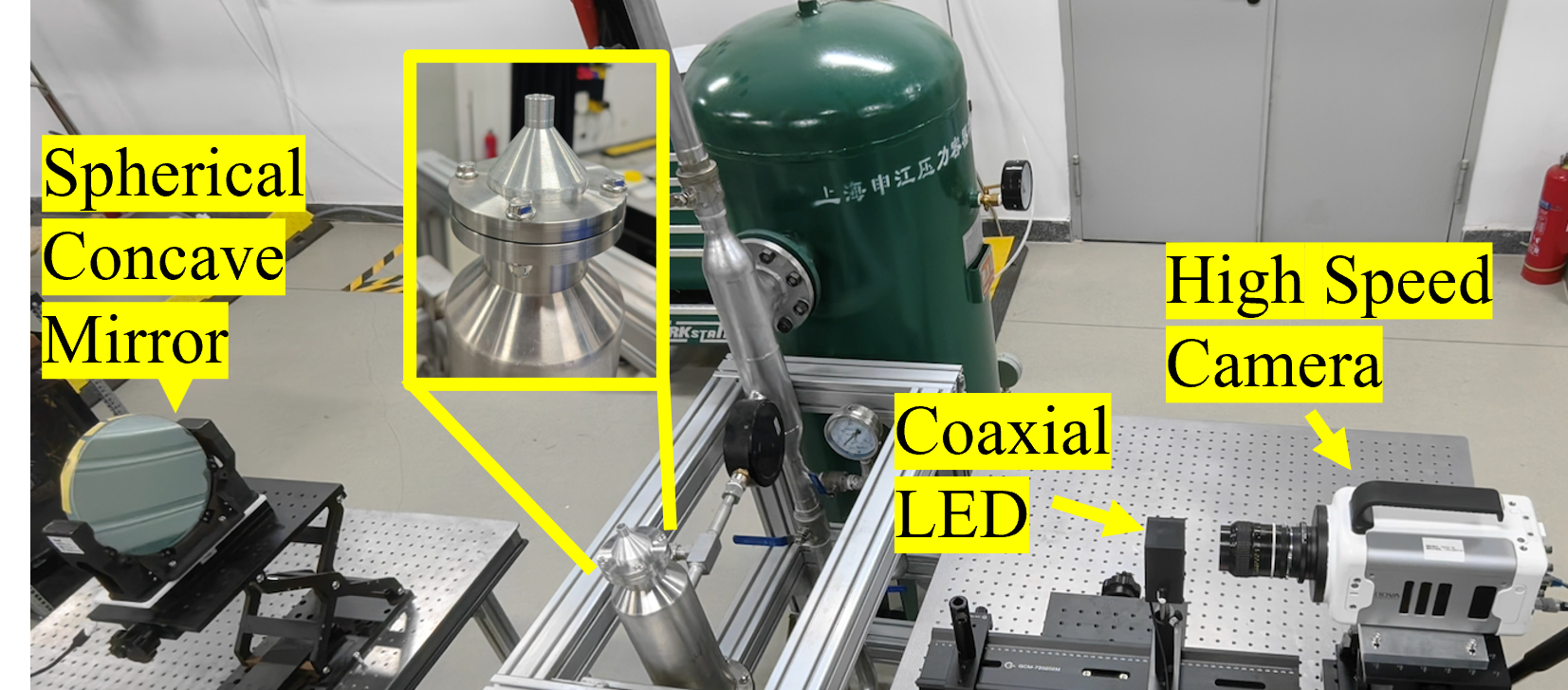}
\caption{Experimental setup for DRBOS and BOS with the supersonic underexpanded air jets.}
\label{fig:DRBOS_supersonic_schematic}       
\end{figure}

Furthermore, a supersonic underexpanded air jet experiment is also conducted, within which the shock waves and expansion waves are responsible for the spatial variations in density. Details of the supersonic nozzle design can be found in the work of \cite{wu_flow_2019}. The total temperature is 300 K, and the total pressure is approximately 5 bar. The overall measurement setup of DRBOS for the supersonic jet is shown in Fig.\ref{fig:DRBOS_supersonic_schematic} and the associated parameters are in Table \ref{tab:underexpanded jet}. For both the sub- and supersonic flow experiments, both BOS and DRBOS are conducted with the same system parameters for comparison purposes. To switch from DRBOS to the canonical BOS, a retroreflective background printed with a similar randomly dotted pattern is inserted in front of the SCM.

\begin{table}[!htb]
\caption{System parameters for DRBOS and BOS with the supersonic underexpanded air jet.}\label{tab:underexpanded jet}%
\begin{tabular}{@{}llll@{}}
\toprule
Parameter & Setting \\
\midrule
SCM diameter(DRBOS) &  $203$ mm \\
Nozzle internal diameter &  $12.7$ mm \\
$f_{\mathrm{m}}$ & $750$ mm \\
$Z_{B}$ &  $1500$ mm \\
$Z_{a}$ & 750 mm \\
Image resolution & $384 \times 336$ pixels\\
Pixel size & $20 \times 20$ $\mu$m$^2$\\
Frame rate & $40$ KHz\\
Exposure time & $25$ $\mu$s\\
$f_{\mathrm{Lens}}$ & $105$ mm \\
$f_{\#}$ & BOS - $22$, DRBOS - $2.8$ \\
\botrule
\end{tabular}
\end{table}

\section{Results and Discussion}\label{sec:results}

\subsection{Spatial Resolution and Sensitivity}\label{sec:DRBOS_Res_Sens}

\begin{figure}[!htb]
  \centering
  \includegraphics[width=71mm]{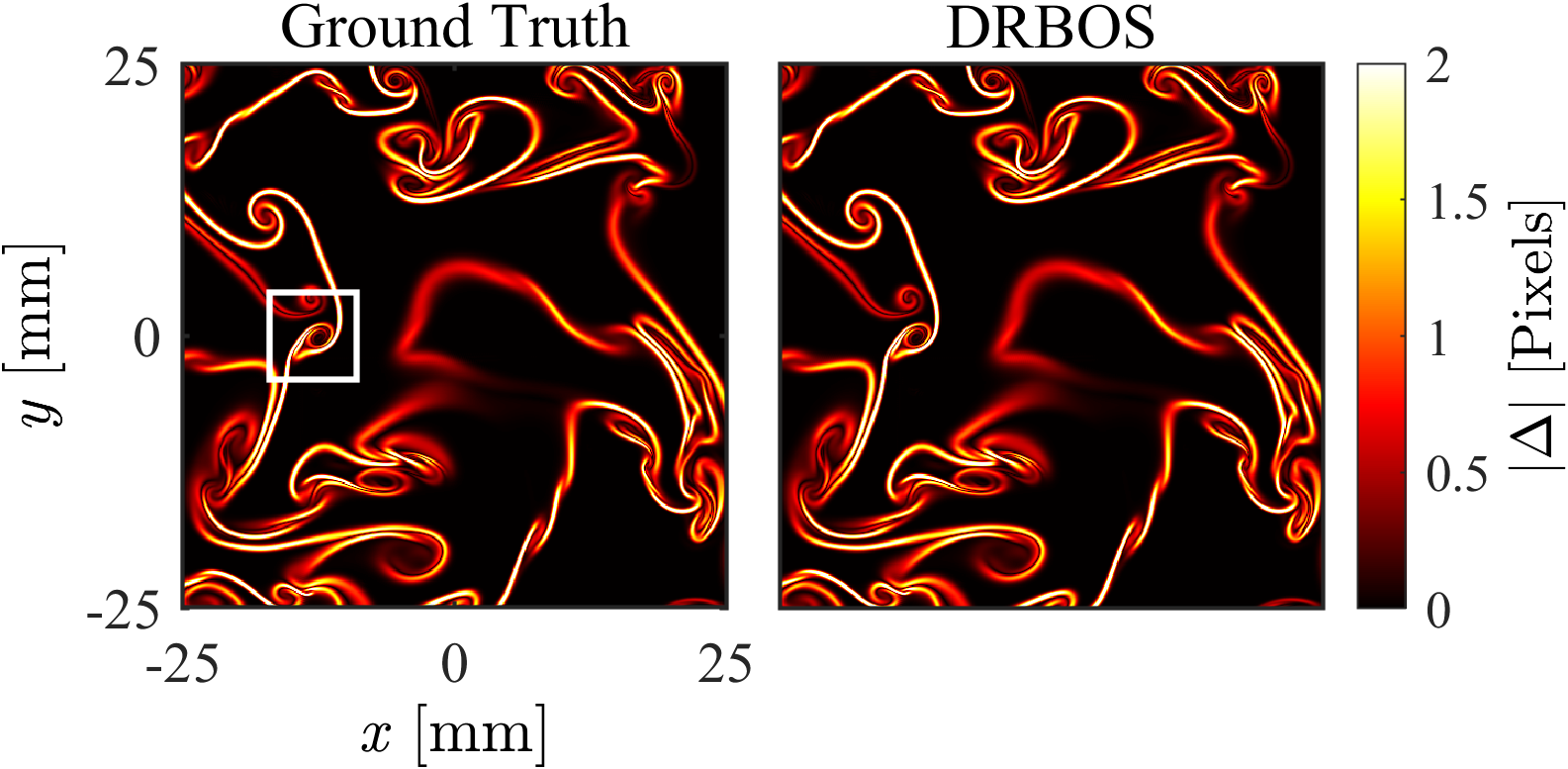}
\caption{Displacement fields obtained with DNS and DRBOS with synthetic DRBOS images.}
\label{fig:SP_DRBOS_synthetic}       
\end{figure}

The capability of the DRBOS to achieve improved spatial resolution is first explored based on synthetic DRBOS images. To simplify the analysis, assume each point on the SCM emits a single ray towards the optical center of the camera lens. With the flow field in Fig.\ref{fig:syn_BOS_fhash}, the extracted displacements are shown in Fig. \ref{fig:SP_DRBOS_synthetic}. Nearly identical displacement fields can be identified between the DRBOS and the ground truth, confirming the superior spatial resolution of DRBOS. This is physically sound since directional rays eliminate the light cones formed due to the combination effect of diffusive background reflections and of a finite-size aperture.

\begin{figure}[!htb]
  \centering
  \includegraphics[width=69mm]{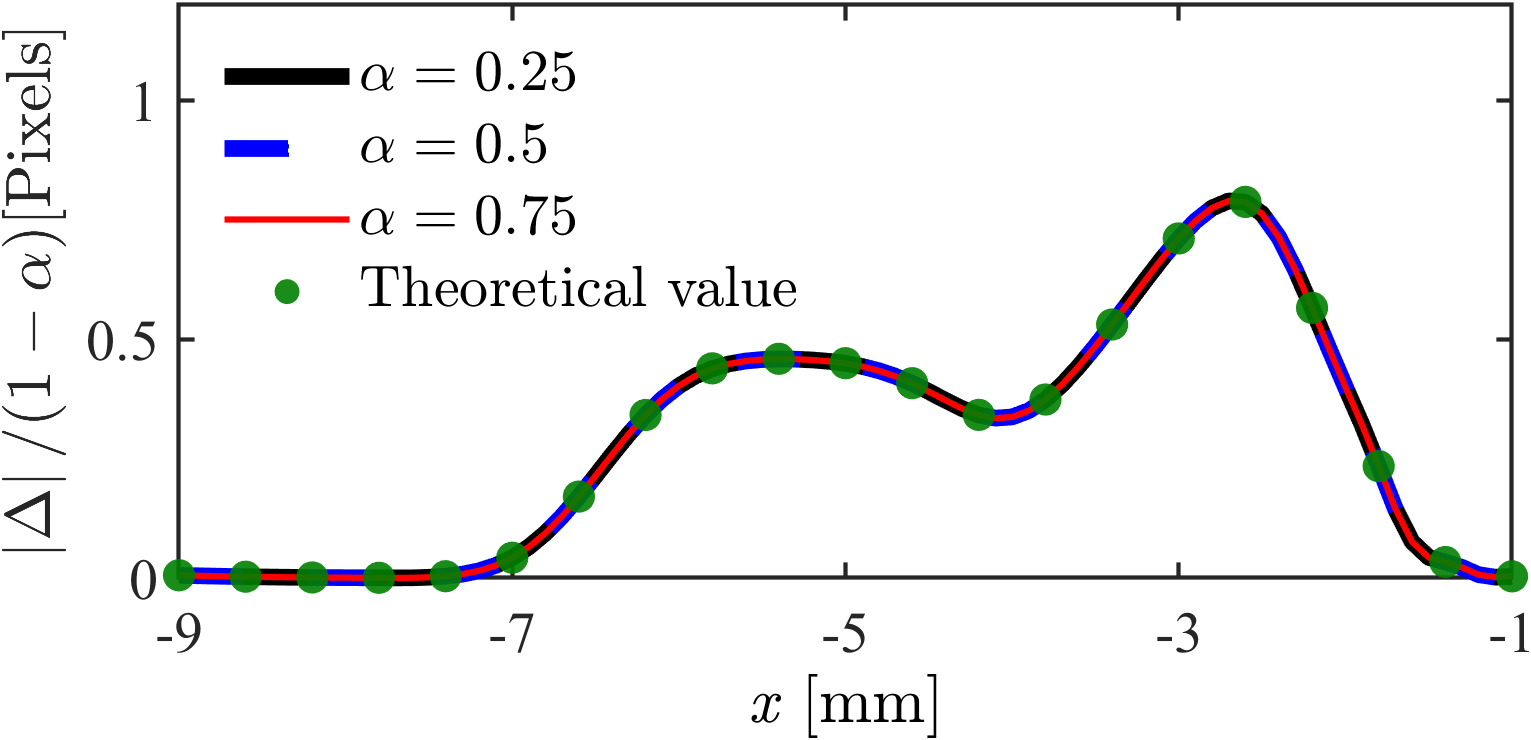}
\caption{Validation of the sensitivity derivation for DRBOS with synthetic DRBOS image generation and theoretically predicted displacements.}
\label{fig:DRBOS_sensitivity}       
\end{figure}

Note that for canonical BOS, improved spatial resolution can also be achieved by allocating the phase object closer to the background at the expense of the measurement sensitivity $S$, which is defined as the displacement distance on the sensor plane $\Delta$ with respect to a certain deflection angle $\varepsilon$: 
\begin{equation}
S=\frac{\Delta}{\varepsilon}=MZ_d=f_{Lens}(1-\alpha).
\label{eq:BOS_sensitivity}
\end{equation}
Inside $\alpha = Z_a/Z_B$ is the location ratio of the phase object between the background and the camera lens and $f_{Lens}$ is the focal length of the camera lens. To verify the superiority of DRBOS compared to the canonical BOS, DRBOS must achieve a high spatial resolution without compromising the measurement sensitivity. According to the sensitivity definition in Eq.\ref{eq:BOS_sensitivity}, DRBOS has $S={(x_4-x_3)}/{\varepsilon}$ according to Fig. \ref{fig:BOS schematic}(b). For a directional ray, we have $x_3 = x_1\times M$ with $M = Z_i/Z_B$. By focusing the camera on the SCM,  $x_4 = x_1^\prime\times M$. Further assume a small deflection angle that $\varepsilon \simeq (x_1^\prime -x_1)/Z_d$, the sensitivity of DRBOS can be obtained, which is identical to Eq.\ref{eq:BOS_sensitivity}. This result indicates that DRBOS can decouple the constraint relationship between spatial resolution and the measurement sensitivity. To validate the sensitivity derivation for DRBOS, the sensitivity value for DRBOS is modified by allocating the flow field at different $\alpha = 0.25, 0.5$, and $0.75$, respectively, for the synthetic DRBOS image generation. As shown in Fig. \ref{fig:DRBOS_sensitivity}, the normalized displacements $\Delta/f_{Lens}(1-\alpha)$ with respect to the normalized $x$ all collapse together with the theoretical displacements calculated according to the formula proposed by \cite{rajendran_pivbos_2019}, confirming the validity of the above derivation. 

\subsection{Coaxial Configuration for DRBOS}\label{sec:DRBOS_dualpath_error}
\begin{figure}[!htb]
  \centering
  \includegraphics[width=71mm]{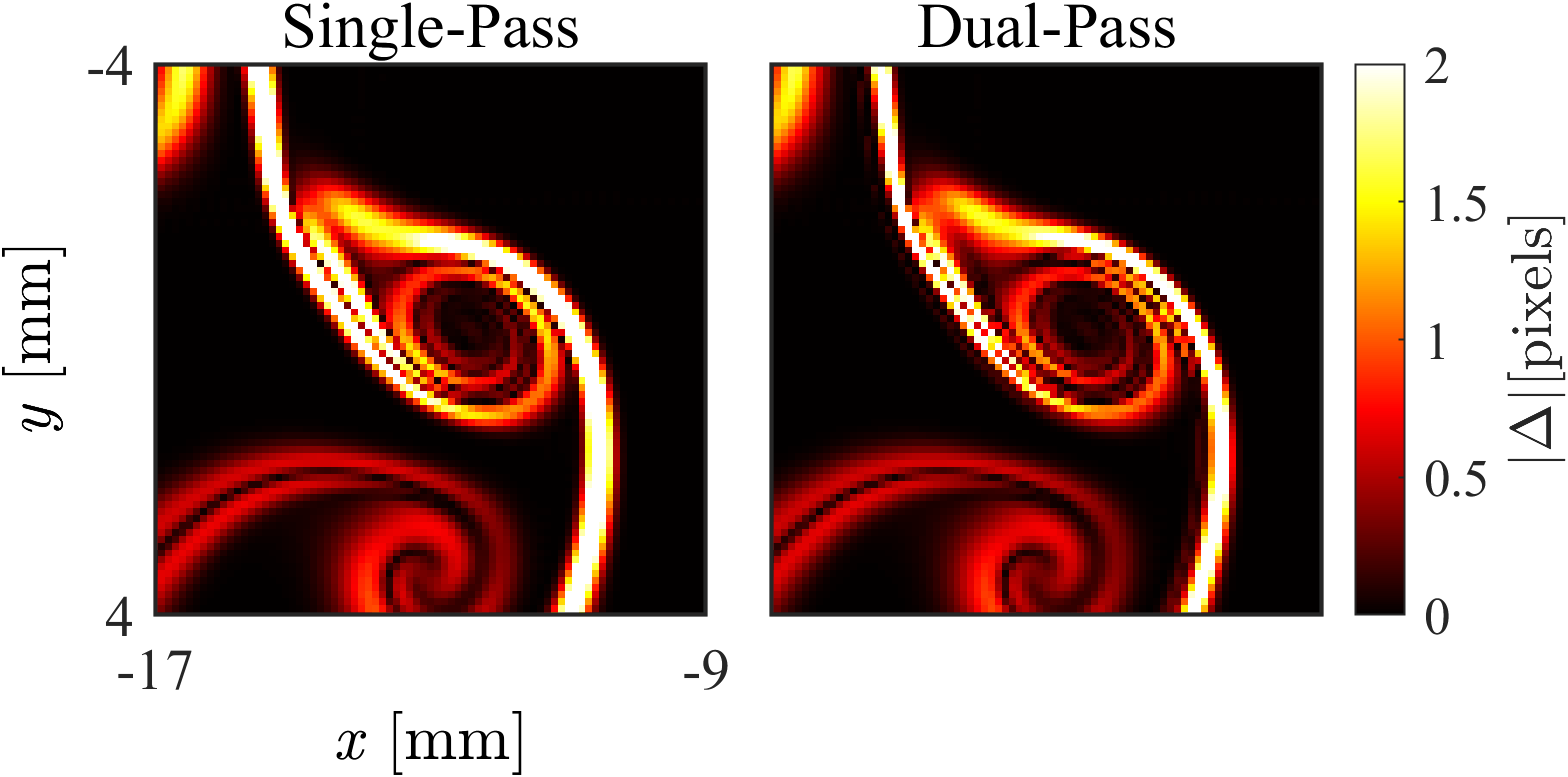}
\caption{A local vortical density structure obtained by assuming 'emitting' SCM and SCM with the coaxial configuration.}
\label{fig:dual_pass}       
\end{figure}

The above analysis assumes the SCM 'emits' rays directly towards the optical center of the lens. However, with the coaxial DRBOS configuration, rays emitted by the LED at $2f_m$ pass through the phase object before arriving the SCM and then get reflected towards the phase object until arriving at the image plane. To reveal the error caused by assuming the 'emitting SCM', synthetic DRBOS images are generated with the coaxial configuration, which precisely allows the light rays to pass through the phase object twice before arriving at the image plane. We note that the rays pointing to the SCM deviate slightly after passing the phase object, and the SCM further magnifies such deviations during the reflection. Consequently, the rays from the SCM are not pointing towards the lens's optical center precisely. In general, extracted DRBOS displacement fields with the coaxial configuration are very similar to the one shown in Fig. \ref{fig:SP_DRBOS_synthetic}, thus not duplicated here with full scale. Instead, a fine vortical structure within the white box region, as marked in Fig.\ref{fig:SP_DRBOS_synthetic}, is shown in Fig. \ref{fig:dual_pass} with and without the coaxial configuration. A slight decrease in the displacement amplitude and width can be clearly identified, especially near the sharp density gradients. This can be explained by the fact that the rays from the LED passing through the thin density gradient will not return to the exact location after SCM's reflection, thus causing broadening and weakening effects. Nevertheless, we will verify experimentally in the following section that the coaxial configuration will introduce only limited errors during the DRBOS measurement, especially for a thin phase object. 

\subsection{Subsonic Helium Jet Experiments}\label{sec:he_jet_exp}
\begin{figure}[!htb]
  \centering
  \includegraphics[width=67mm]{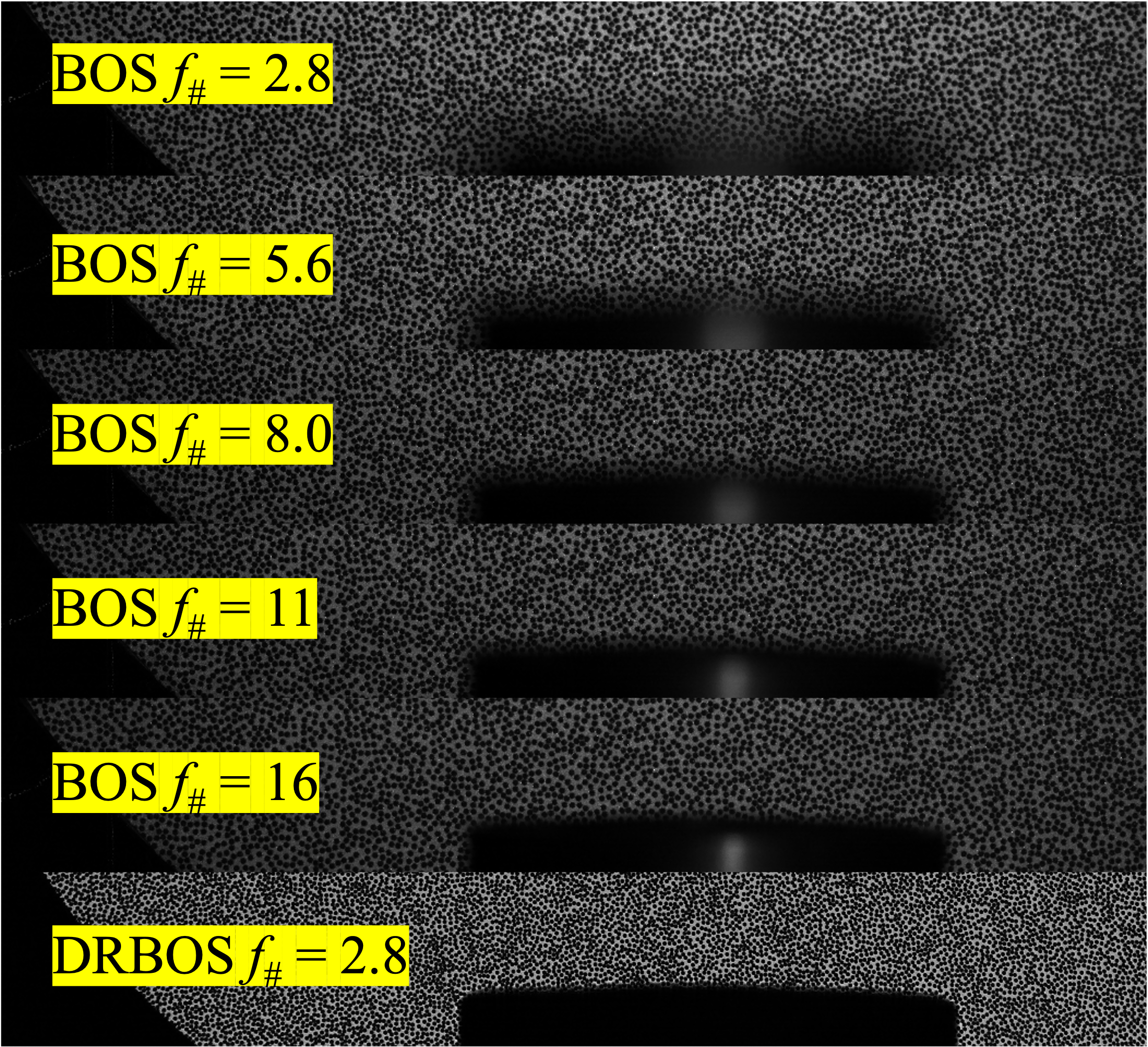}
\caption{Raw images of the jet nozzle for the canonical BOS and DRBOS measurements with different $f_\#$.}
\label{fig:raw_nozzle}       
\end{figure}

With the Helium jet configuration in Sec. \ref{sec:sub_expsetup}, canonical BOS measurements are firstly implemented with the aperture size varying from $f_\# = 2.8$ to $16$. Since the out-of-focus bluriness is proportional to $f_\#$, we can first check the blurriness of the raw image as shown in Fig. \ref{fig:raw_nozzle}. For $f_\#$ varying from 2.8 to 16, the sharpness of the jet nozzle increases, consistent with the increased depth-of-field. With DRBOS, even with $f_\# = 2.8$, a sharp shadowgraph of the jet nozzle can be clearly identified with a more precise boundary contour, even compared to the BOS case with $f_\# = 16$. Such a comparison indicates that the flow field reconstructed by DRBOS should also be less blurred compared to the BOS with $f_\# = 16$. 

\begin{figure}[!htb]
  \centering
  \includegraphics[width=122mm]{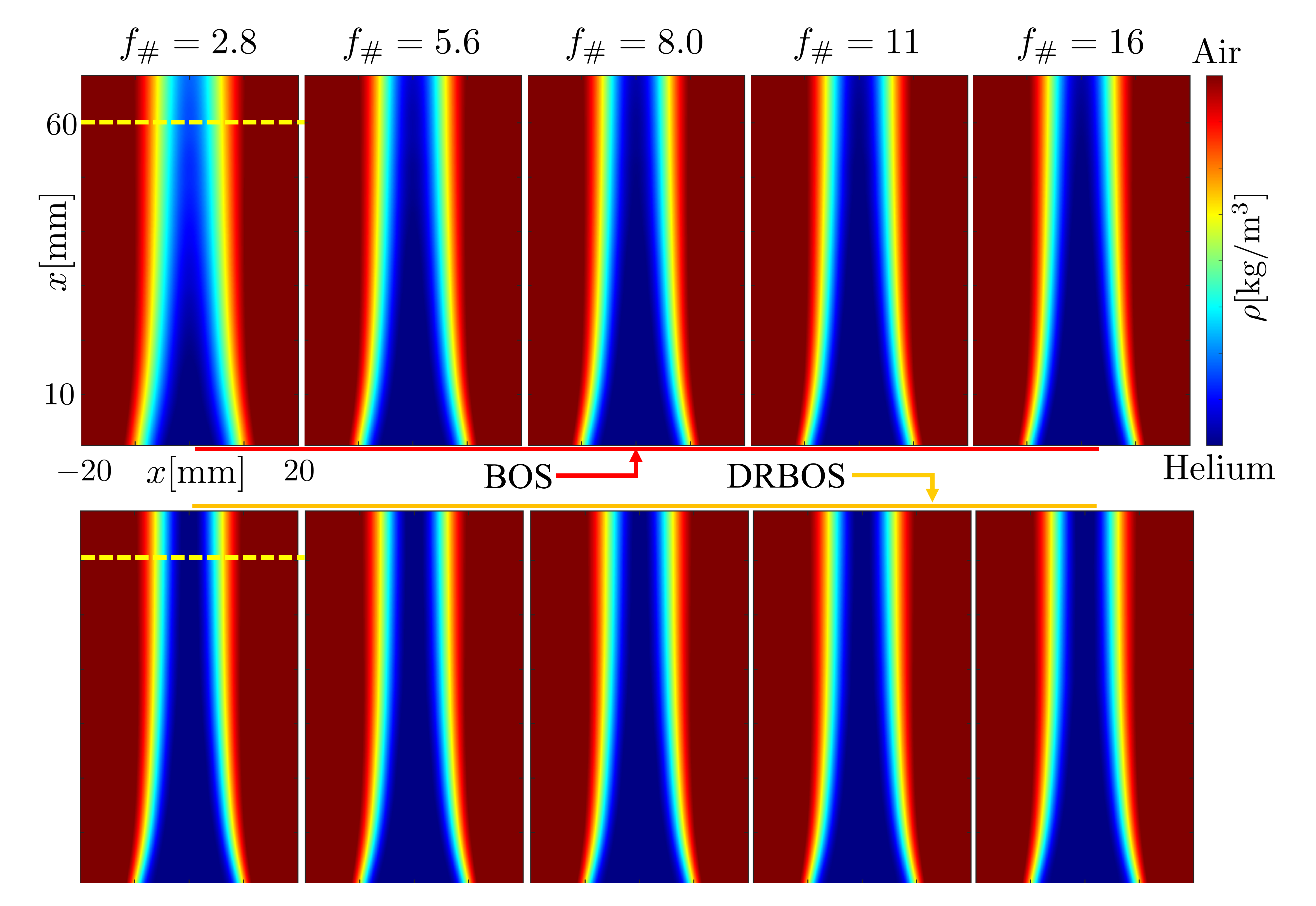}
\caption{Comparison of the density fields reconstructed by BOS and DRBOS, respectively, with varying $f_\#$ from 2.8 to 16.}
\label{fig:hejet_canonical_synthetic}       
\end{figure}

Reconstructed density fields for the Helium jet at different $f_\#$, using both the canonical BOS and DRBOS, are shown in Fig.\ref{fig:hejet_canonical_synthetic}. Regarding the canonical BOS measurements, the blurriness of the jet density field, especially downstream of the jet, becomes sharper as $f_\#$ increases, consistent with the observations on the raw nozzle images shown in Fig. \ref{fig:raw_nozzle}. For comparison purposes, we repeat the same experiments with coaxial DRBOS, varying the aperture size in the same range. Reconstructed density fields are shown in the bottom row of Fig.\ref{fig:hejet_canonical_synthetic}. Firstly, we can note that DRBOS reconstructs density fields that are much sharper compared to those measured with BOS, especially at $f_\# = 2.8$. Moreover, those density fields measured by DRBOS are nearly identical irrespective of $f_\#$. This unique advantage of DRBOS encourages the application of DRBOS in high-speed flow applications, which usually require an ultra-strong background illumination with the canonical BOS with a small aperture size. 

\begin{figure}[!htb]
  \centering
  \includegraphics[width=122mm]{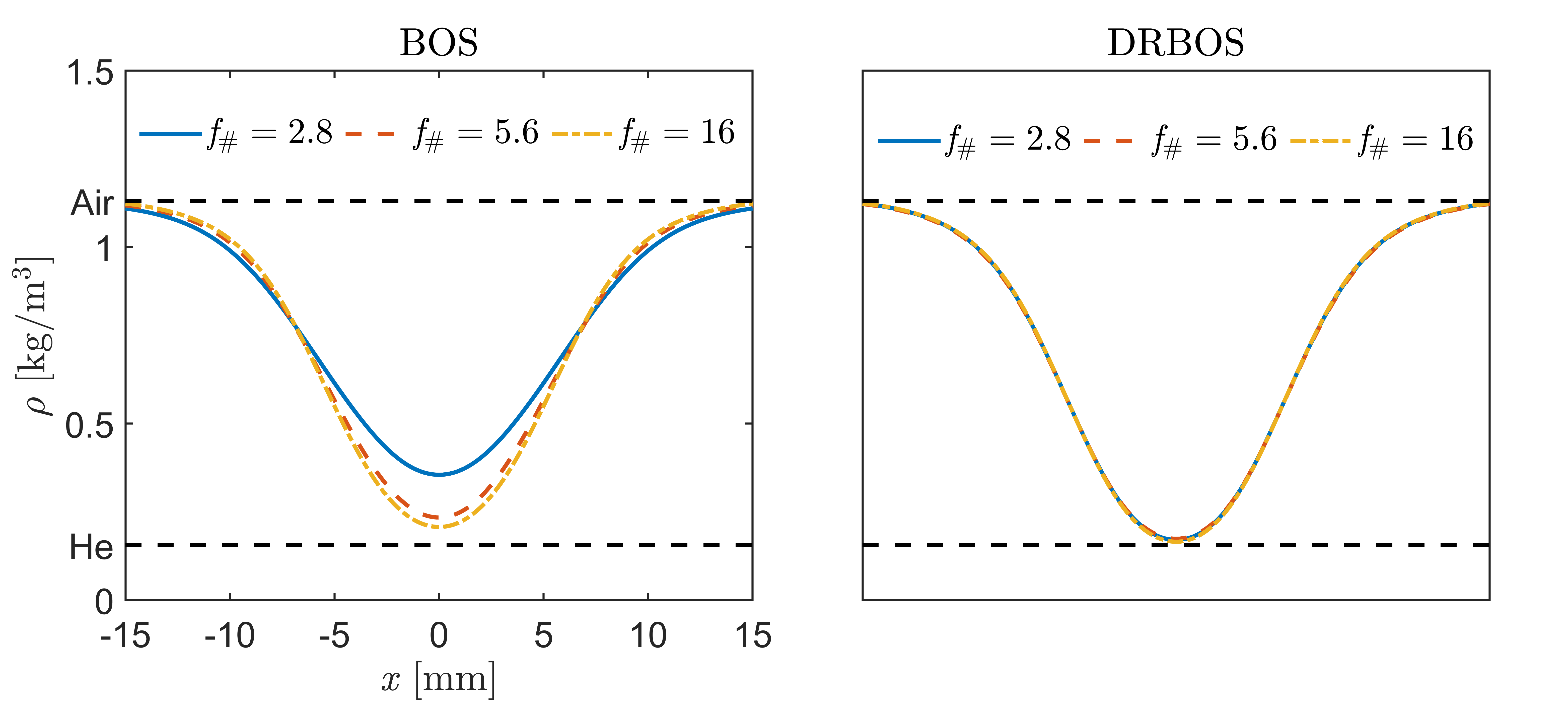}
\caption{Quantitative comparison of the density profile obtained with BOS and DRBOS at $y = 60 \rm mm$.}
\label{fig:fhash_quantitative}       
\end{figure}

Quantitative density profiles at $h = 60 \rm mm$ downstream of the jet exit are further shown in Fig. \ref{fig:fhash_quantitative}. For the BOS measurements, the peak lower density is approaching Helium density as $f_\#$ increases. However, a slight difference still exists between the two, even with $f_\# = 16$. This deviation indicates that reducing the aperture size can sometimes be insufficient to reach the desired spatial resolution possibly due to the diffraction contribution to the blurred image. Comparatively, regardless of the lens aperture, DRBOS reconstructed density profiles at different $f_\#$ all collapse. The lowest density value approaches the Helium density better compared to that of BOS at $f_\# = 16$. Note that the Helium jet is allocated roughly in the middle between the background and the camera, maximizing the measurement sensitivity \cite{lang_measurement_2017}. This setting indicates that DRBOS can reach the maximized measurement sensitivity without compromising the spatial resolution.

\subsection{Supersonic Jet Experiments}\label{sec:supersonic_jet_exp}
\begin{figure}[!htb]
  \centering
  \includegraphics[width=67mm]{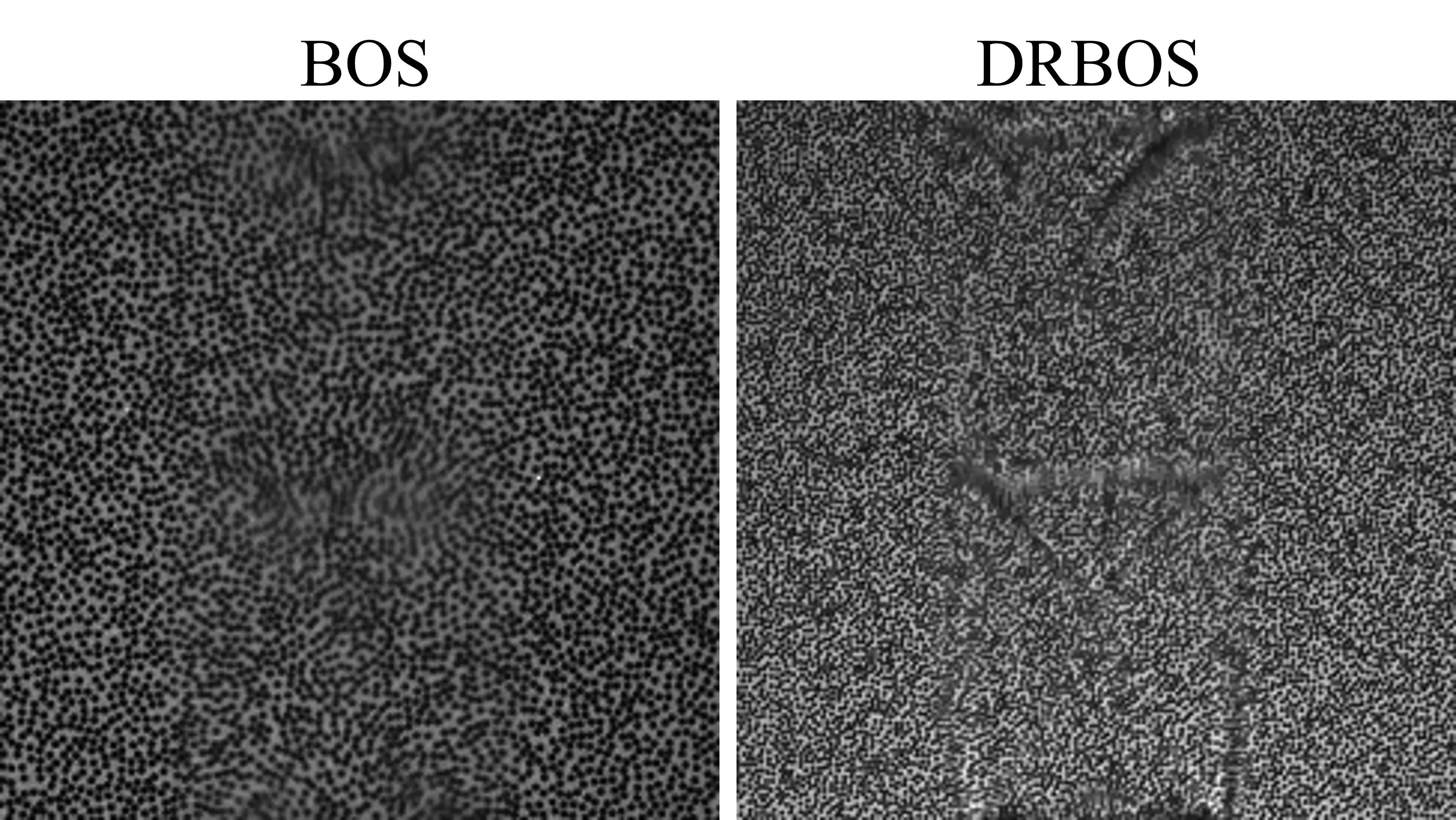}
\caption{Background raw images with the supersonic jet on captured with BOS and DRBOS, respectively.}
\label{fig:shock_raw}       
\end{figure}

As demonstrated in the Helium jet experiments, DRBOS can work with no loss in the LED light energy owing to the SCM's focusing nature. Thus, applying the DRBOS for high-speed flows at a very high framing rate becomes feasible even with a cheap LED. To demonstrate the high-framing-rate performance of DRBOS, we conduct the supersonic underexpanded air jet experiments as detailed in Section \ref{sec:sub_expsetup}. Capturing the supersonic density structure within such jets is challenging for the canonical BOS to reach a high spatial resolution. A raw image of the background with the supersonic jet is shown in Fig. \ref{fig:shock_raw} captured with both BOS and DRBOS. The spatial resolution limits of BOS in capturing the jet structure are obviously illustrated. Note that for the BOS measurements, the camera lens aperture $f_\#$ is already set to 22, reaching the limit for most commercially available lenses.
Nevertheless, the supersonic shock structure is completely blurred due to the compromise in the spatial resolution. As a comparison, the DRBOS raw image exhibits a typical shock wave structure, indicating its strong spatial resolving capability. Note that both images within Fig.  \ref{fig:shock_raw} are captured at a 40 kHz framing rate. For BOS with $f_\# = 22$, an 18W high-power LED (HARDSOFT) is utilized together with a retroreflective background to boost the light illumination. For comparison, DRBOS uses an LED with only 0.3W to achieve the same brightness. 

\begin{figure}[!htb]
  \centering
  \includegraphics[width=120mm]{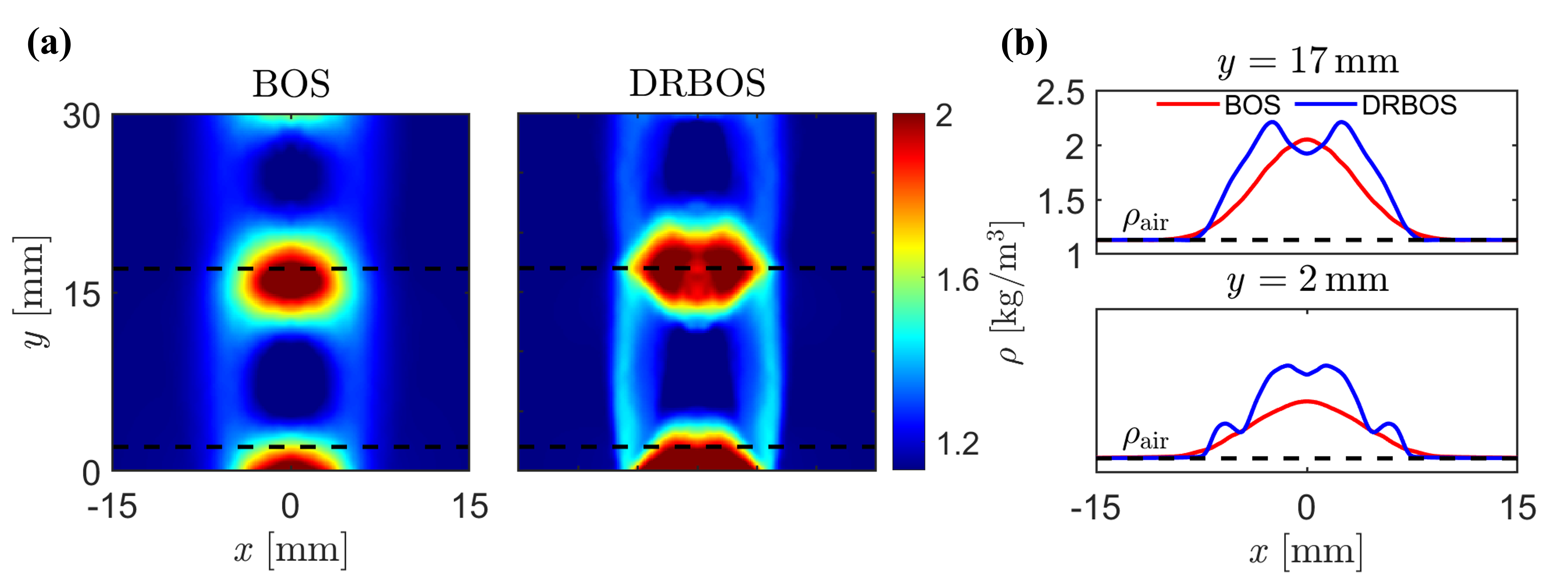}
\caption{Density fields reconstructed with BOS and DRBOS at 40 kHz framing rate.}
\label{fig:shock_wofa}       
\end{figure}

Extracting the displacement field using the wavelet optical flow algorithm by \cite{schmidt_wavelet-based_2021} and following the indirect axisymmetric density flow reconstruction by \cite{xiong_towards_2020}, the density structure of the supersonic jet can be obtained as shown in Fig. \ref{fig:shock_wofa}(a) for both BOS and DRBOS. Similar to the observations on the raw image, the shock structure is completely blurred and nearly invisible for BOS, and the central compressed pocket tends to be ellipsoidal. For comparison, DRBOS reconstructs the sharp density interface and the diamond-like density structure typically visible in supersonic underexpanded jets. A quantitative comparison of the density profiles at two different heights marked as dashed lines within Fig. \ref{fig:shock_wofa}(a) is shown in Fig. \ref{fig:shock_wofa}(b). While canonical BOS yields only a Gaussian-like density profile with the smallest aperture size, DRBOS can reveal complex density ripples along the radial direction, confirming the superiority of DRBOS in terms of the spatial resolution.

\section{Conclusions}\label{sec:Conclusions}
The DRBOS technique proposed in this paper fundamentally addresses the spatial resolution limitation inherent to the canonical BOS without compromising the measurement sensitivity. By employing a spherical concave mirror with an etched random pattern of dots, nominal directional rays are generated, eliminating the cone-averaging effect caused by diffusive background reflection in the canonical BOS configurations. Theoretical analysis confirms that DRBOS maintains the same sensitivity as the canonical BOS while achieving substantially improved spatial resolution. Comprehensive validation through both synthetic image analysis and experimental measurements demonstrates the superiority of the DRBOS in terms of high spatial resolution and low requirements for background illumination. DRBOS with coaxial configuration introduces minor systematic errors in predicting the displacement amplitude and width, but can gain significant simplifications in terms of the optical setup, thus should be encouraged for practical applications, especially for thin phase objects. Future applications of DRBOS appear particularly promising for wind tunnel testing and other scenarios where conventional BOS struggles with limited spatial resolution and illumination constraints.

\bmhead{Acknowledgements}
This work is supported by the National Natural Science Foundation of China (NSFC Grant No.12572320, No.12502329), the National Key Research and Development Program of China (2024YFA1612300), and the Fundamental Research Funds for the Central Universities.

\bmhead{Data availability}
Data underlying the results presented in this paper are not publicly available at this time but can be obtained from the authors upon reasonable request.






\bibliography{references}

\end{document}